\documentclass[prd,aps,preprint,epsfig,floats]{revtex4}
\input epsf.tex
\def\DESepsf(#1 width #2){\epsfxsize=#2 \epsfbox{#1}}
\begin{document}

%\draft
\preprint{\vbox{
\hbox{UMD-PP-03-062} }}

\title{\Large\bf Lepton Electric Dipole Moments, Supersymmetric Seesaw
 and Leptogenesis Phase}
\author{\bf Bhaskar Dutta$^1$ and R.N. Mohapatra$^2$ }

\affiliation{$^1$ Department of Physics, University of Regina, Regina, SK,
S4S, Canada\\$^2$ Department of Physics, University of Maryland, College Park,
MD 20742, USA}

\date{July, 2003}
%\maketitle

\begin{abstract}
We calculate the lepton electric dipole moments in a class of
supersymmetric seesaw models and explore 
the possibility that they may provide a way to probe some of the CP
violating phases
responsible for the origin of matter via leptogenesis. We show that in
models where the right handed neutrino
masses,  $M_R$ arise from the breaking of local B-L by a Higgs field with
B-L=2, some of the
leptogenesis phases can lead to enhancement of the lepton dipole
moments compared to the prediction of models where
$M_R$ is either directly put in by hand or is a consequence of a higher
dimensional operator.
\end{abstract}
\maketitle
\section{Introduction}
    A presently popular way to understand the origin of matter-anti-matter
asymmetry in the universe is to start with a mechanism to generate lepton
asymmetry in the early Universe using the CP violating effects in the
decay of a heavy right handed Majorana neutrinos\cite{fy} and let the
sphaleron interactions\cite{krs} above the electroweak phase transtion
temperature convert them to a baryon asymmetry. This is possible since the
sphaleron interactions which violate B+L symmetry are in thermal
equilibrium\cite{krs} above the eletcroweak phase transition
temperature. The right handed neutrino decay is of course
not the only way to generate the pre-electroweak lepton asymmetry.
Several other ways have been discovered for the purpose
(e.g. see references \cite{sm,others}). We will
however focus in this paper on the decay of heavy right handed neutrinos
since they also provide a simple way to understand small
neutrino masses via the seesaw mechanism\cite{seesaw}. In particular,
the existing experimental information on neutrino oscillations via the
seesaw mechanism effects the nature of the right handed neutrino mass
pattern. which in turn effects the magnitude of the baryon asymmetry.
It is quite interesting that detailed analyses that use the exisitng 
neutrino data do indeed give the right magnitude for baryon
asymmetry as well as important insight into the pattern of right 
handed neutrino masses\cite{buch}.

As noted in the original pioneering work of Sakharov, CP violation is
an essential ingredient in the generation of any particle-anti-particle 
asymmetry. In the present case, the CP violating decays of the right
handed neutrinos arise
from the phases in right handed neutrino couplings and we will call them
leptogenesis phases. It is clearly important to seek low energy
manifestations of
the leptogenesis phase for several reasons: first, this will improve our
understanding of the right handed neutrino mass
matrix which plays a crucial role in the neutrino mass physics
at low energies; secondly, it may shed light on the origin of the seesaw
mechanism, which will then provide a useful window into physics beyond the
standard model. Moreover, since the seesaw formula
provides some connection between the low and high scale phases in the
theory, understanding leptogenesis phases may be a guide to the CP
violating effects in neutrino oscillations, with its many experimental
ramifications.

There have been a great deal of discussion of this issue in
literature \cite{davidson}. Specific models where the neutrino mass phase
and leptogenesis phases are directly related have also been discussed in
several papers\cite{endoh} providing one way to probe the latter. However
in all previous discussions of this issue, the masses of the right handed
neutrinos $M_R$ are either put in by hand or are assumed to arise from
nonrenormalizable operators. It is however well known that in
models that contain Higgs bosons $\Delta$ with B-L=2, $M_R$ arises from a
renormalizable coupling $f \nu^c\nu^c \Delta$. Examples of such theories
are left-right symmetric models with triplet Higgs fields or
SO(10) models with {\bf 126} dimensional Higgs fields, $SU(5)_L\times
SU(5)_R$ models with Higgs fields in $(1,15)\oplus (15,1)$
representation. In such theories, there are new
renormalization group running effects between the GUT (or Planck) scale
and the seesaw scale. We showed in a previous paper\cite{babu} that the
presence of these renormalizable couplings can lead to quantitatively
different effects in seesaw induced lepton flavor violation. In
particular, the ratio $B(\mu\rightarrow e+\gamma)/B(\tau\rightarrow
\mu+\gamma)$ depends on the SUSY breaking parameters in very different
ways in different theories.

In this paper, we calculate the electric  dipole moments (edm) in the
class of supersymmetric seesaw models with the $f$ coupling and discuss
how it helps in probing the leptogenesis phase.

Since we are working within supersymmetric seesaw models, (i) we need to
make assumptions about the nature of supersymmetry breaking and (ii) the
embedding of MSSM into new physics at the seesaw scale. 

As for the first point, in common
with many discussions in the literature,
we will assume that the TeV scale theory is the minimal supersymmetric
standard model, which solves the gauge hierarchy problem  plus the
additional feature that the neutrinos have
Majorana masses arising from the seesaw mechanism. As far as SUSY
breaking is concerned, we will
work within the context of minimal MSUGRA models\cite{msugra} and
assume that around the GUT or the Planck scale, all soft SUSY breaking
scalar masses are
universal as are the gaugino masses and the A-terms are proportional to
the Yukawa couplings. Furthermore we will assume that there are no
overall phase
in the $A$ terms, which is guaranteed if for example, MSSM is part of a
left-right
symmetric SUSY model near the GUT scale\cite{rasin}. The resulting theory
is a very economical one with
only five parameters characterizing the complete susy breaking sector of
 MSSM and is therefore quite predictive. More importantly,
the restrictive nature of this assumption means that at low energies, we
are only tracking
the effect of the phases in the right handed neutrino couplings
responsible for leptogenesis and no other phase unrelated to neutrino
physics is present. Once we give up these assumptions, new phases could
come in and confuse the issue.

As far as the high theory theory is concerned,
to implement our
premise of having seesaw mechanism arise out of a B-L = 2 Higgs field,
we will consider the high scale theory to be  $SU(2)_L\times
U(1)_{I_{3R}}\times U(1)_{B-L}$ based with a standard model singlet
$\Delta$ with B-L =2 breaking the B-L gauge symmetry and also giving the
seesaw mechanism. This model can arise as an effective theory from
left-right or SO(10) theories. This extended model
 leaves the low energy predictions of MSSM uneffected.

 The weak scale values of the supersymmetry breaking
parameters can then be derived by the renormalization group extrapolations
using the standard techniques. Since it is the dimension four terms in the
Lagrangian which are largely responsible for the running, how the right
handed neutrino masses are generated in the Lagrangian does make
difference to low energy phenomenology. We discuss the impact of this
extra running effect on the lepton dipole moment observable which depends
on the leptogenesis phases.

The main result of our investigation is
that if the right handed neutrino mass in the seesaw mechanism
arises from the vev of a B-L=2 Higgs field, it has the effect of
enhancing the lepton electric dipole moments over models where the
right handed neutrino mass is put in by hand. We give two examples to
illustrate this point: one where the neutrino mixings arise
purely from
Dirac Yukawa coupling of the right handed neutrinos and a second one where
it arises from their Majorana coupling $f$. We also discuss the case
of a $3\times 2$ seesaw models. In all the cases, a B-L=2 Higgs field is
responsible for righthanded neutrino masses.

 We present the results of our calculation 
for electron amd muon edms for the above models making sure that we stay
in the range of parameters which fit all neutrino data and generate 
the right amount of lepton asymmetry. We find that leptogenesis phase can
 lead to enhanced edms for leptons
in the presence of the $f$ couplings than in their absence (i.e. where the
RH neutrino has a bare mass term.), although the effects are still
small. Most optimistic values for electron edms in our models are 
 between $10^{-30}$ to $10^{-31}$ ecm and for muon edm between
$10^{-27}$ to $10^{-28}$ ecm.  
It is encouraging that there are ideas
and plans for drastic improvement in the search for the lepton edms in the
future\cite{lamo,seme,miller}. For instance, search for electrons edm upto
the level of $10^{-31}$ ecm\cite{lamo} and muon edm to $10^{-26}$ ecm
have been contemplated\cite{seme}.

 Although the values predicted by our
analysis are small, high precision searches for lepton edms to the
contemplated level can still teach us something useful. For instance, if
the edms discovered are above our predictions, it will mean one of several
things: either there is gross departures from the assumed forms for
supersymmetry breaking terms or if independent experiments such as at
LHC or Tevatron have confirmed the SUSY breaking assumptions (universality
and
proportionality), we would have to conclude that baryogenesis does not
originate via leptogenesis. Either of these conclusions would be very
important in our search for physics beyond the standard model.

We have organized this paper as follows: in sec. 2, we present a generic
model based on the gauge group $SU(2)_L\times
U(1)_{I_{3R}}\times U(1)_{B-L}$ to discuss the general calculational
framework and the basic idea of our work; in sec 3, we discuss the
renormalization group effects of the neutrino sector on the supersymmetry
parameters. in sec.4, we present
a qualitative explanation of how the enhancement of edm can arise in our
approach; in sec. 5, we briefly discuss the calculation of baryogenesis; 
in sec. 6 we discuss the first neutrino mass model, give
the neutrino mass fits, calculation of lepton asymmetry and present the
results for lepton edms; in sec. 7 and 8, we repeat the same discussion
for two other models; in sec. 9, we summarize our results and present
the conclusions.

\section{A generic model and the calculational framework}

To proceed with our discussion, let us write the superpotential for our
model, which must be invariant under the gauge group $SU(2)_L\times
U(1)_{I_{3R}}\times U(1)_{B-L}$:
\begin{eqnarray}
W~=~e^{cT}{\bf Y_{\ell}}L H_d+ {\nu^c}^T{\bf Y_{\nu}}L H_u+ {\bf
\frac{f}{2}\Delta}{\nu^c}^T\nu^c + S(\bar{\Delta}\Delta+\lambda H_uH_d+S^2
- v^2_R)
\end{eqnarray}
Here $ L, e^c, \nu^c$ are leptonic superfields; $H_{u,d},\Delta,
\bar{\Delta}$ are the Higgs fields. We do not display the quark part of
the superpotential but it is same as in the MSSM.
 
The first point to notice is that it is a very simple extension of the
MSSM superpotential, that leads to a vacuum with
 ${\bf <\Delta>}=<\bar{\Delta}>=v_R$. This leads to the RH neutrino mass
matrix $M_R~=~f v_R$ and clearly it
arises from a renormalizable term in the Lagrangian.

This superpotential has also another advantage that it solves the $\mu$
problem\cite{babu1} i.e. it predicts that $\mu\simeq m_{3/2}$,
where $m_{3/2}$ is the superpartner mass, which is of order of a
TeV. To see this, notice that this potential
has an R-symmetry under which $W\rightarrow -W$; (the fields transform as
$(H_u,e^c,d^c)\rightarrow
i(H_u,e^c,d^c)$; $(H_d, \nu^c,u^c)\rightarrow
-i(H_d,\nu^c,u^c)$; $S\rightarrow -S$
and all other fields remain unchanged under R-symmetry). The vanishing of
the F-terms then
imply that $<S>=0$ and $<S>$ arises from supersymmetry breaking terms. As
a result $\mu\simeq <S>\sim m_{3/2}$.

The seesaw formula (type I) can be written as:
\begin{eqnarray}
{\cal M}_{\nu}~=~-{\bf Y^T_\nu f^{-1}Y_{\nu}}
\frac{v^2_{wk}tan^2\beta}{v_R}
\label{seesaw}\end{eqnarray}
We can choose a basis where the coupling matrix $f$ is diagonal as is the
charged lepton Yukawa coupling matrix $Y_\ell$. The neutrino Dirac
coupling matrix $Y_\nu$ is then a $3\times 3$ complex matrix whose
diagonal elements can be made real by suitable redefinition of the phases
of the $\ell_i$ and $e^c_i$ fields. This leaves us with nine real
parameters in the $Y_\nu$ matrix and six phases. We rewrite the $Y_\nu$
matrix as follows:
\begin{eqnarray}
Y_\nu~=~V^* Y^d_\nu W^{\dagger}
\label{wv}\end{eqnarray}
where $V~=~P_1\tilde{V}P_2$ with
$P_i~=~diag(e^{i\phi_{i1}},e^{i\phi_{i2},1}$ and $\tilde{V}$ and $W$ are
CKM type matrices with only one phase in each. Thus all six phases are
accounted for.

To see which of the phases are responsible for lepton asymmetry, let us
write down the formula for lepton asymmetry, $n_\ell$\cite{buch1}:
\begin{eqnarray}
n_\ell\propto \sum_j Im [Y_\nu Y^{\dagger}_\nu]^2_{1j} F(\frac{M_1}{M_j})
\label{nl}\end{eqnarray}
where symbol 1 in the above equation denotes the lightest right handed
neutrino eigenstate.
Note that using Eq.\ref{wv}, one can conclude that only the phase in
$\tilde{V}$ and those in $P_1$ contribute to lepton asymmetry. The phases
in $P_2$ completely drop out.  Our goal will therefore be to see how we
can measure the phases in $P_1$ and $\tilde{V}$ by low energy experiments.

Secondly, we find two results for the three generation case that we state
below.

\subsection{Independence of light neutrino mixings and leptogenesis
phases}

 The first point is that the lepton asymmetry parameter
$n_\ell$ is independent of the PMNS mixing angles. To show this first note
that
\begin{eqnarray}
{\cal M}_\nu~=~U^*{\cal M}^d_\nu U^{\dagger}~=~~-{\bf Y^T_\nu
f^{-1}Y_{\nu}}m_0
\end{eqnarray}
where $m_0~=~\frac{v^2_{wk}tan^2\beta}{v_R}$ and $U$ is the PMNS
mixing matrix.
We can now invert the seesaw relation to conclude that
\begin{eqnarray}
Y_\nu~=~if^{1/2}O({\cal M}^d_\nu)^{1/2}U^{\dagger}
\label{ynu}\end{eqnarray}
where $O$ is a complex matrix with the property that
$OO^T=1$\cite{casas}. The set of
matrices $O$ in fact form a group analogous to the complex
extension of the Lorentz group. Using Eq.\ref{ynu} in Eq.\ref{nl}, we see
that lepton asymmetry $n_\ell$ is independent of the PMNS mixing angles in
$U$\cite{rebelo}. This result is important because what this
means is that if there are three right handed neutrinos, low energy
experiments such as neutrino oscillations and neutrinoless double beta
decay give us no information about the leptogenesis phase.

\subsection{Vanishing of lepton asymmetry for degenerate neutrino masses}  
To see this, we rewrite the formula for $n_\ell$ in a slightly
different way.
Using the seesaw formula in Eq.\ref{seesaw} and assuming that there is a
mass hierarchy among the right handed neutrinos, we can write\cite{buch2}
\begin{eqnarray}
n_\ell \propto Im [Y_\nu {\cal M}^{\dagger}_\nu Y^T]_{11}
\label{buch}\end{eqnarray}
Using this and Eq.\ref{ynu}, we can write
\begin{eqnarray}
n_\ell \propto Im[f^{1/2}O({\cal M}^d_\nu)^2 O^Tf^{1/2}
\label{nl1}\end{eqnarray}
From this we can conclude that when the neutrino masses are degenerate
i.e.
${\cal M}^d_\nu = m{\bf I}$, then $n_\ell =0$ since $f_{11}$ is real. This
result, to the best of
our knowledge does not exist in literature. A similar result for the
manifestly CP
violating observable $P_{\nu_e-\bar{\nu}_e}-P_{\bar{\nu}_e-\nu_e}$ was
noticed in Ref.\cite{andre}, where it was pointed out that it vanishes for 
degenerate neutrinos.

 An implication of this result
is that if neutrinoless double beta decay results imply $m\geq 0.1$ eV(see
e.g. \cite{klap}), then the
lepton asymmetry from RH neutrino decay will be suppressed by an extra
factor of 100 and we might then have to look for extra enhancements or
completely different way of obtaining the baryon asymmetry of the
universe.

\section{RGE's and low energy effects of leptogenesis phases}
Let us now proceed to the main part of our discussion, which involves the
supersymmetry breaking superpartner proiperties at low energies.

It is clear that one simple way to measure high scale parameters of a
theory is to look for low energy observables that owe their origin or
magnitude to the
renormalization group evolution (RGE) from the high scale to the low scale
(i.e. without RGE effects, the low energy theory will predict the
observable to
be very tiny.) In our case such observables are lepton flavor violating
transitions as well as lepton electric dipole moments (edms). In the
absence of the RGE effects, we expect $B(\mu\rightarrow e+\gamma)$ to be
proportional to $\frac{m^4_\nu}{m^4_W}\sim 10^{-48}$
As far as the edm of leptons go, in the low energy theory without any
other effects
i.e. standard model with a Majorana
mass for the neutrino, we would
expect $d_e\simeq \frac{G_Fm_e}{16\pi^2}\frac{m^2_\nu}{m^2_W}$ to be of
order $10^{-48}$ ecm. 

In supersymmetric models, the presence of superpartners introduce new
contributions to both LFV as well as edm observables at the one loop
level, where virtual superpartners flow inside the loop. The resulting
LFV and edm effects are therefore not only enhanced but they are also
sensitive to the nature of SUSY breaking terms. In this paper we will
assume
a very restrictive form for the SUSY breaking terms which defines the so
called mSUGRA models. According to this assumption, the scalar masses
and all gaugino masses are
universal at the GUT or the Planck scale; the $A$ terms are proportional
to the Yukawa couplings.  These assumptions are motivated in
several SUSY breaking scenarios; they not only reduce the number of
parameters in the theory but may be necessary for understanding the
observed suppression of flavor changing effects in both quark and lepton
sectors. Furthermore, we will assume that the overall phase in the $A$
term is zero. This can happen for example, if the theory is part of SUSY
left-right model at the GUT scale\cite{rasin}. This is necessary in order
to resolve the so-called SUSY CP problem.

The proportionality assumption coupled with the assumtion of no overall
phase in the $A$ term guarantees that all the CP
violating phases in the lepton sector sector are in the $Y_\nu$ and
therefore any CP violating low energy effect after renormalization group
evolution is only sensitive to the phases that are manifested in the
leptogenesis as well as the neutrino oscillations via the seesaw
mechanism. 

As mentioned, we need to extrapolate the SUSY breaking parameters from the
GUT or Planck scale down
to the seesaw scale where we will perform the calculation of the electric
dipole moments of the electron and the muon. This extrapolation will
always involve the couplings $Y_\nu$ and the diagonal coupling matrix
$Y_\ell$.  Extrapolation below the seesaw scale only involves the $Y_\ell$
which is mostly diagonal to leading order and does not effect any of our
results. We will keep only one loop effects. The relevant renormalization
group equations are:
\begin{eqnarray}
\frac{dY_{\nu}}{dt}~=~\frac{Y_{\nu}}{16\pi^2}
[Tr(3Y_uY^{\dagger}_u+Y_{\nu}Y^{\dagger}_{\nu}) +3Y^{\dagger}_{\nu}Y_{\nu}
 +Y^{\dagger}_{\ell}Y_{\ell} -3g^2_2-
g^2_{R}-3/2g^2_{B-L}]+\frac{1}{16\pi^2}ff^{\dagger}Y_{\nu}\\
\frac{dm^2_L}{dt}~=~\frac{1}{16\pi^2}[ (m^2_L + 2m^2_{H_d})
Y^{\dagger}_\ell Y_\ell
+(m^2_L+2H^2_u)Y^{\dagger}_{\nu}Y_{\nu} +2Y^{\dagger}_\ell m^2_{e^c}Y_\ell
+Y^{\dagger}_\ell Y_\ell m^2_L\\ \nonumber+2Y^{\dagger}_\nu
m^2_{\nu^c}Y_\nu
+Y^{\dagger}_\nu Y_\nu m^2_L
 +2A^{\dagger}_\ell A_\ell +2A^{\dagger}_\nu A_\nu
 -6g^2_2M_2^2-3g^2_{B-L}M_{B-L}^2]
\end{eqnarray}
Let us also write down the renormalization group equations for the
$A$ parameters- specifically the $A_{\ell}$.
\begin{eqnarray}
\frac{dA_\ell}{dt}&=& \frac{1}{16\pi^2} [A_\ell [ Tr (3Y^{\dagger}_dY_d +
Y^{\dagger}_\ell Y_\ell) + 5 Y^{\dagger}_\ell Y_\ell+Y^{\dagger}_\nu Y_\nu
\\ \nonumber
&-&3g^2_2-g^2_{R}-3/2 g^2_{B-L}] +Y_\ell[Tr(6A_dY^{\dagger}_d+2 A_\ell
Y^{\dagger}_\ell)\\ \nonumber &+& 4 Y^{\dagger}_\ell A_\ell+
2Y^{\dagger}_\nu A_\nu+
6g^2_2M_2 +2g^2_{R}M_{R}+3g^2_{B-L}M_{B-L}]]
\end{eqnarray}
\begin{eqnarray}
\frac{dA_\nu}{dt}&=& \frac{1}{16\pi^2} [A_\nu [ Tr (3Y^{\dagger}_uY_u +
Y^{\dagger}_\nu Y_\nu) + 5 Y^{\dagger}_\nu Y_\nu+Y^{\dagger}_\ell Y_\ell
\\ \nonumber
&+& -3g^2_2-g^2_{R}-3/2g^2_{B-L}+ 4ff^{\dagger}A_{\nu}]
+Y_\nu[Tr(6A_uY^{\dagger}_u+2 A_\nu Y^{\dagger}_\nu) \\ \nonumber&+&
4 Y^{\dagger}_\nu A_\nu+2 Y^{\dagger}_\ell
A_\ell+6g^2_2M_2 +2g^2_{R}M_{R}+3g^2_{B-L}M_{B-L}]+8 A_ff^{\dagger}Y_\nu]
\end{eqnarray}
While in presenting actual results, we numerically integrate these RGEs to
obtain their contribution to physical observables such as lepton edms, it
is instructive to give some approximate analytic expressions for the
different results to get feeling for our final predictions. We do this in
the following section.

\section{Estimates of the electron and muon edms and connections to
leptogenesis phase}
To see the new effects from the $f$ couplings in a qualitative manner, we
note that
the dominant supersymmetric contributions to the lepton dipole moments
come from the one loop diagram involving the bino. A rough estimate of
this is
\begin{eqnarray}
d^e_{\ell_i}\simeq \frac{g^2_1}{8\pi^2 M^4_{\tilde{\ell},0}}
Im (A_{\ell}M^2_{\ell})
\end{eqnarray}
If there is an overall phase in $A_\ell$, then one can easily estimate
$d^e_e\simeq
10^{-23} sin\delta$ ecm. To be consistent with present limits i.e. $\leq
1.7\times 10^{-27}$ ecm, one must have phase $\delta \leq 10^{-3}$. This
is the so called SUSY phase problem. We assume that in the theory the SUSY
CP problem has been solved, (e.g. by embedding the MSSM into a SUSY LR
model as mentioned) so that we can set $\delta =0$. In this limit,
 in the absence of the RGE effects, since
 $M^2$ is real, in this limit, all edms vanish. If this term dominated the
contribution to the lepton dipole moments, then since the $A$ term id
proportional to charged lepton masses, we would have the scaling law
$d^e_\mu/d^e_e=m_\mu/m_e$. In any theory where this law holds, the present
upper limit on electron edm of $\leq 4.3\times 10^{-27}$ ecm would imply
that $d^e_\mu \leq 8.6\times 10^{-15}$ ecm. Also this scaling law is a
signature of the simplicity of the underlying theory. In the
supersymmetric context, there exist exceptions to this scaling
law\cite{babu2}.

After RGE effects are taken into account, lepton edms receive
contributions from $Y^{\dagger}_\nu Y_\nu$ which is not a diagonal matrix
as well as $f^{\dagger}f$ and $Y^2_\ell$ which are diagonal matrices at
the GUT scale. The $M^2$ and $A_\ell$ at the seesaw scale are polynomials
in these matrices and in order to get a nonvanishing contribution to
$d^e_\ell$, we must look for imaqginary diagonal elements in the product
$A_\ell M^2_{\tilde{\ell}}$, which will involve products of
$Y^{\dagger}_\nu Y_\nu$, $f^{\dagger}f$ and $Y^2_\ell$.

The leading order term in $A_\ell M^2_{\tilde{\ell}}$ in the absence of
the $f$ couplings (i.e. a bare
mass term for $\nu_R$), which has a complex $11$ entry is of  the form
\begin{eqnarray}
d^e_\ell \propto Im[m_0Y_\ell Y^{\dagger}_\nu Y_\nu Y^2_\ell
(Y^{\dagger}_\nu Y_\nu)^2].
\label{simple}\end{eqnarray}
From Eq. (3), it is easy to see that $Y^{\dagger}_\nu Y_\nu$ is
independent of the matrix $V$ that is responsible for
leptogenesis. One might conclude from this that simple RGE extrapolations
like this term will never enable us to probe the matrix $V$. However, as
has been argued in\cite{sacha}, since $Y^{\dagger}_\nu Y_\nu$ is
sensitive to the unitary matrix $W$ in Eq. (3). So in principle, one can
get $W$ from the experimental information about supersymmetry breaking
sector. Then, if one is
given the right handed neutrino mass spectrum and the form of $Y_\nu$, one
 can get all the matrix elements of $V$ using the relation
\begin{eqnarray}
V^TfV~=~-(Y^d_\nu)^{-1}W^TU^*{\cal M}^d_\nu U^{\dagger}W(Y^d_\nu)^{-1}
\end{eqnarray}
and hence the
baryogenesis phases. This would require knowledge of all the matrix
elements of slepton mass matrix. While this is possible in principle, one
would like a more ``experimentally feasible'' way to probe the phases in
$V$. This is what we pursue here.

There are two ways in which the RGE extrapolation can still leave traces
of the matrix $V$. One way discussed in \cite{ellis} is the following.
In the case where the right handed neutrino masses are hierarchical, the
simple RGE expression in the above paragraph gets modified by threshold
effects as follows: the simple expression in Eq.\ref{simple} remains valid
above the highest right handed neutrino mass, $M_3$; once we go below this
i.e. for $M_2\leq \mu\leq M_3$, only the second and the lightest RH
neutrinos contribute. This means that in the matrix $Y_\nu$, only
$Y_{\nu,2i}$ and $Y_{\nu,1i}$ contribute and below $M_2$ only $Y_{\nu,1i}$
entries contribute. Therefore traces of the matrix $V$ remain in the low
energy expressions. To get a large effect, a
significant gap between the $M_3$ and $M_1$ is required.

Another way that $V$ can be visible is the presence of the
$f$ terms, which give the leading contribution to edm to be
$Y^{\dagger}_\nu Y_\nu Y^{\dagger}_\nu f^{\dagger}fY_\nu$ instead of 
what is given in \ref{simple}. Thus in this case, there is a higher order
effect even in the absence of threshold effects.

 Note further that
each time a term of the form $Y^{\dagger}_\nu Y_\nu$ or $f^{\dagger}f$
appears, there is a loop suppression factor $\epsilon_{loop}\simeq
\frac{1}{16\pi^2}\ell n\left(\frac{M_U}{M_R}\right)\simeq
0.06-0.1$. Taking this into account we see that the second contribution
(i.e. one in the presence of the $f$ couplings) is enhance by the
inverse of one loop factor compared to the first one. Also, there is an
additional effect coming from the gap between the $M_U$ or $M_{P\ell}$ and
$M_3$. Therefore barring unexpected cancellations, we expect the edm
contribution in the presence of $f$ contributions to be bigger. In our
numerical integrations to present the final numbers for edms, we keep the
threshold effects.

\section{Overview of Baryogenesis estimate}

As already mentioned earlier, we use the mSUGRA framework to calculate the
electric dipole moments of the
 muon and the electron. The value of the universal scalar mass $m_0$,
 universal gaugino mass $m_{1/2}$, universal trilinear term $A_0$,
$\tan\beta$ and  the sign of $mu$ as free parameters determine our
final result. We also assume that there is no phase
 associated with the SUSY breaking. The Yukawa and/or the Majorana
couplings are responsible for  CP violation in these models.
 
The mSUGRA parameter space is constrained  by the experimental
lower limit on $m_h$ and the measurements of $b\rightarrow s\gamma$ and the
recent results on dark matter relic density\cite{wmap}. For low
$\tan\beta$, the parameter space has lower bound on  $m_{1/2}$ from the
light Higgs mass bound of $m_h\geq 114$ GeV. For larger $\tan\beta$ the
lower bound on $m_{1/2}$ is produced by the CLEO constraint on the
BR($b\rightarrow s\gamma$). The recent WMAP results have led to stronger
constraints on dark matter density which in turn reduces the allowed
parameter space of mSUGRA mostly to the co-annihilation region for
$m_0,\,m_{1/2}\leq 1000$ GeV.

We  calculate the baryon to photon ratio $\eta_B$ in three different models
for neutrino masses. The SUSY breaking parameters do not play a role in
this discussion. The lepton asymmetry is produced from the out of
equilibrium decays of right handed neutrinos and this 
gets converted into baryon asymmetry through the sphaleron processes. The
formula for $\eta_B$ is:
\begin{equation}
\eta_B={n_B\over{n_{\gamma}}}={a g_s^0}\sum{{\kappa_i \epsilon_i}\over{g^*}}
\end{equation}
where $a$ relates the B-L asymmetry to baryon asymmetry and 
${a g_s^0\over{g^*}}\sim 5\times 10^{-3}$. $\kappa$ is the washout factor 
\cite{buch1} and
$\epsilon_i={1\over {8\pi}}\sum_{k\neq
i}(f({{|M_k|^2}\over{|M_i|^2}})+g({{|M_k|^2}\over{|M_i|^2}})){{Im[(h^{\dag}h)^2_{ik}]}\over{(h^{\dag}
h)_{ii}}}$\cite{buch1} with
$f(x)=\sqrt{x}[ln({{1+x}\over{x}})]$ and 
$g(x)=2{\sqrt{x}\over{1-x}}$.

The recent experimental bound on $\eta_B$ is\cite{wm}:
\begin{equation}
\eta_B=(6.5\pm^{0.4}_{0.3})\times 10^{-10}
\end{equation}
In the following sections, we will use these expressions to evaluate the
baryon asymmetry, which along with present neutrino oscillation
results will then give us a range of parameters in the seesaw formula. We
will use these parameters to calculate the edms of the electron and the
muon using the allowed range of SUSY breaking parameters mentioned above.

We present three different cases.
 In at least one case, the effects are
big enough that they are within reach of contemplated experiments
for the elctron\cite{lamo}. Planned searches for the muon edm\cite{miller}
will also throw important light on this issue.

\section{First model for neutrino masses and its predictions for
lepton edms}
In this section, we consider the first of three examples for neutrino
masses and obtain the predictions for lepton edms in the presence of the
$f$ terms. In this example, we take the $Y_\nu$ and $Y_\ell$ to be
diagonal
and assume that all neutrino mixings to arise from arbitrary complex $f$
coupling\cite{babu}. In the notation of sec. 2, this means that $W$ is a
identity
matrix. Clearly therefore, in the absence of the $f$ term contributions,
edm becomes very tiny ($\leq 10^{-48}$ ecm. 
The $f$ couplings lead to an enhancement of the effects of
leptogenesis phases in the lepton edms as we see below. 

As for the $Y_\nu$, we consider two subcases: (a) in the first case
 we let the diagonal elements of $Y_\nu$ be free and (b) in this second
case, we let them be proportional to the charged lepton masses. 

In both cases, we can write the matrix elements of $f$ in terms of the
diagonal elements of $Y_\nu$ i.e. $(c\epsilon^3,\epsilon,1) y$ and the
neutrino mass matrix. For this purpose, we take the light neutrino
mass matrix to be:
\begin{eqnarray}
{\cal M}_\nu = m_0 \left(\matrix{e \epsilon^n & h \epsilon^m & d\epsilon
\cr
h \epsilon^m & 1+a \epsilon & 1 \cr d \epsilon & 1 & 1+b\epsilon}\right)~.
\end{eqnarray}
Here $(a,~b,~d,~e,~h)$ are order one coefficients, $\Delta m^2_{atm}
\simeq 4
m_0^2$ and the exponents $(n,~m)$ in the (1,1) and (1,2) entries have to
be at least 1, but can be larger.
This matrix provides a good
fit to all neutrino data. Using these, we get for $f$ matrix:
\begin{eqnarray}
f={1 \over d^2 m_0} \left(\matrix{(a+b)c^2 \epsilon^5 & cd \epsilon^3 &
-cd
\epsilon^2 \cr cd \epsilon^3 & -d^2 \epsilon^2 & dh \epsilon^2 \cr
-cd \epsilon^2 & dh \epsilon^2 & (e-h^2)\epsilon^2}\right)~.
\end{eqnarray}

We now calculate the edms and $\eta_B$ in this model.
For example, at $\tan\beta=30$:\\
we have ${f}$ at $2\times 10^{12}$ GeV
\begin{eqnarray} {f}&=&\left(\matrix{
  -1.14\times10^{-4}                & -1.56\times 10^{-2}   &0.293
\cr
 -1.56\times 10^{-2}             &0.379- 0.353 i      &-0.21+ 0.55 i \cr
    0.293       &-0.21+ 0.55 i        &0.058-0.091*i   }\right)
\end{eqnarray}
We now assume that the Dirac neutrino mass is proportional to the charged
lepton mass and are
$0.0337, \,6.94, \,123.398$ GeV. Using these parameters, we find
$\sin^22\theta_{\odot}=0.8$, $\Delta m^2_{\odot}=5\times 10^{-5}$ eV$^2$
and
$\sin^22\theta_{A}=0.98$, $\Delta m^2_A=6\times 10^{-3}$ eV$^2$. In the
calculation we decouple
the neutrinos at the respective mass scales.
The baryon to photon ratio is $\eta_B\sim 6\times 10^{-10}$.
We will use these inputs to evaluate the lepton edms. Our results for edms
are given in Fig 1 and 2.

In Fig.1 we plot the electric dipole moments of electron and the muon as
functions of $A_0$ and $m_{1/2}$ and $m_0$ for $\tan\beta=10$. We do not
show explicitly the values
of $m_0$ since we choose their values in such a way so that the relic
density constraint
is satisfied in the only available stau-neutralino co-annihilation region
for the
parameter space. We  apply the recent relic density
constraint i.e. $0.094<\Omega
h^2<0.129$ (2$\sigma$)\cite{wmap}. Using this constraint, the maximum
value of $m_{1/2}$ is found to be 800
GeV for $\tan\beta=10$.
The magnitude $d_{\mu}$ can be as large as $10^{-29}$ ecm for $A_0=800
GeV$, where
as the $d_{e}$ can be as large as $10^{-34}$ ecm. 

In Fig.2, we show the $d_{\mu}$  and the $d_{e}$  for the same model
using $\tan\beta=30$. As expected the edms are larger in this case with
$d_{\mu}$ as large as $4\times 10^{-28}$ ecm.

\begin{figure}\vspace{0cm}
    \begin{center}
    \leavevmode
    \epsfysize=8.0cm
    %\epsffile[75 160 575 630]{edmbdmbl10.eps}
    \epsffile[75 160 575 630]{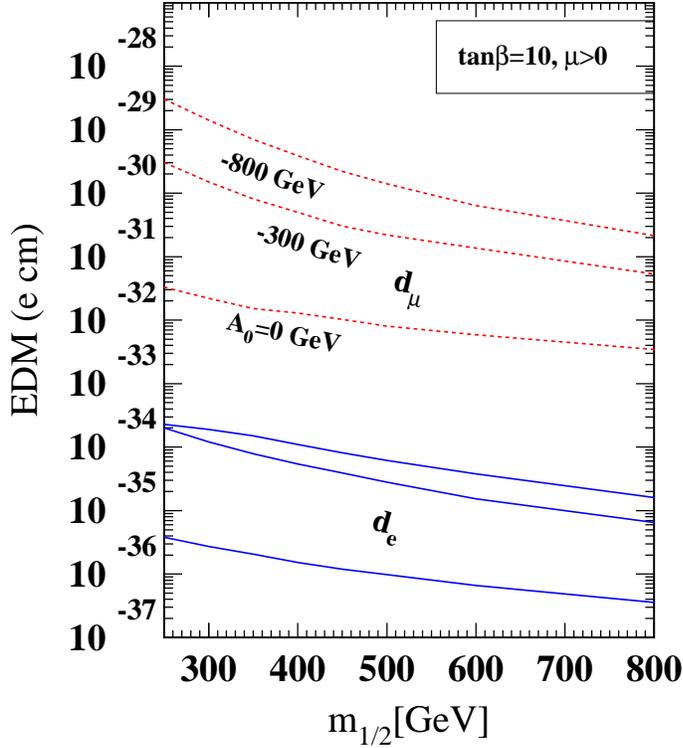}
    \vspace{2.0cm}
     \caption{\label{fig:fig1} Electric dipole moments of electron
$|d_e|$ (solid
     line) and muon $|d_{\mu}|$ (dotted line) for different values of
$A_0$ for
     $\tan\beta=10$.}
\end{center}\end{figure}
\begin{figure}\vspace{0cm}
    \begin{center}
    \leavevmode
    \epsfysize=8.0cm
    %\epsffile[75 160 575 630]{edmbdmbl30.eps}
    \epsffile[75 160 575 630]{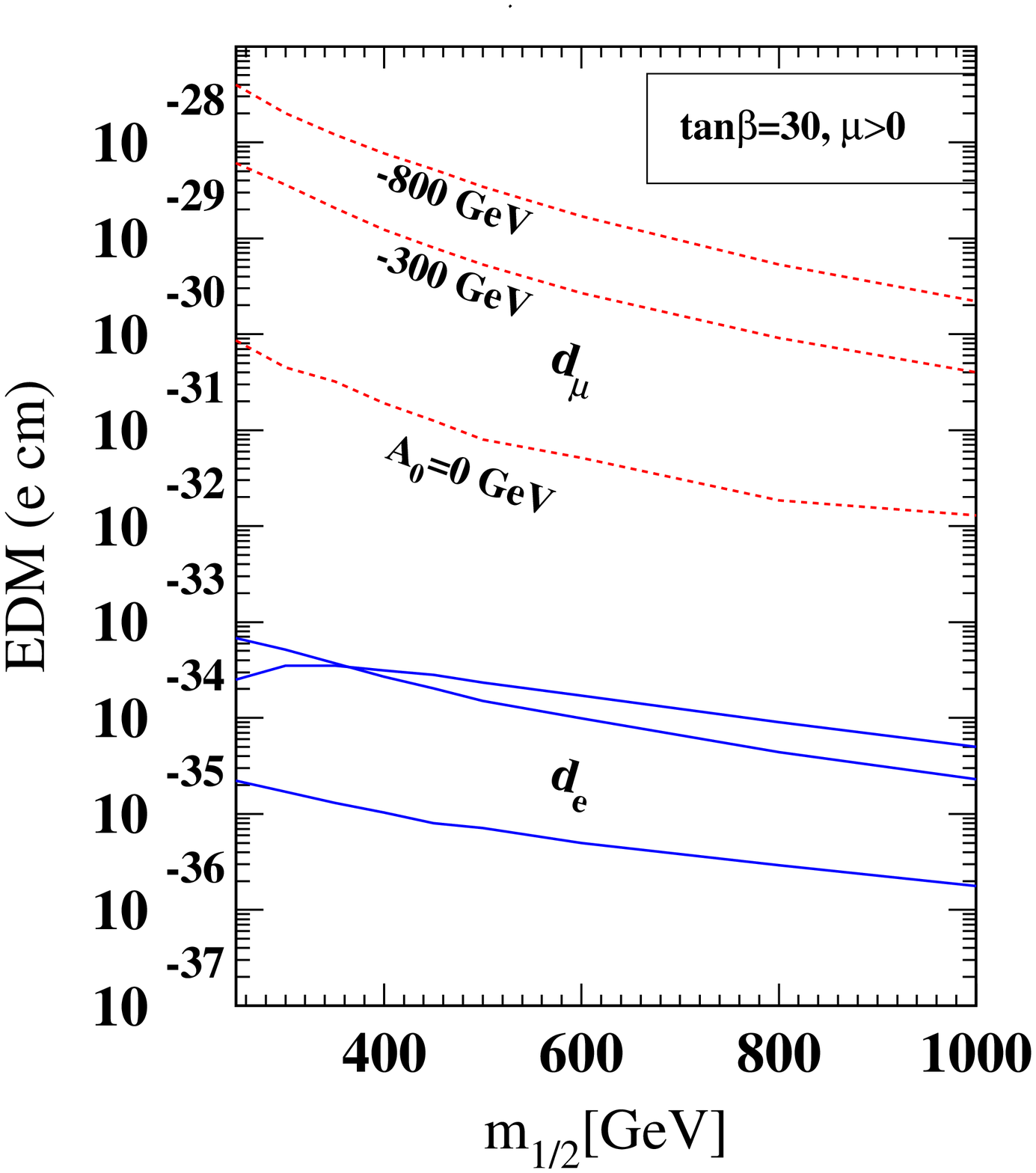}
    %\epsffile[75 160 575 630]{/home/duttabh/lfv/muegmh30.eps}
    \vspace{2.0cm}
     \caption{\label{fig:fig2} Electric dipole moments of electron $d_e$ (solid
     line) and muon $d_{\mu}$ (dotted line) for different values of $A_0$ for
     $\tan\beta=30$.}
\end{center}\end{figure}

\section{Model II for neutrino masses and predictions for lepton edm}
 In this case, we choose $Y_\ell$ and $f$ matrices to be diagonal and $f$
equal to $Diag(\epsilon^6_f, \epsilon^4_f,1)$ but keep
a general form for $Y_\nu$ as follows:
\begin{eqnarray}
{ Y}_\nu =  \left(\matrix{0 & h \epsilon^3 & d\epsilon^3
\cr
h \epsilon^3 & a\epsilon^2 & b\epsilon^2 \cr d \epsilon^3 & b\epsilon^2 &
e^{i\psi}}\right)~.
\end{eqnarray}
For $\epsilon\simeq \epsilon_f\ll 1$ and $a,b\sim 1$ but complex, we get
large solar and
atmospheric neutrino mixings. We have done a detailed fit to the neutrino
parameters keeping $a,b,h,d$ complex and have calculated the baryon
asymmetry and the lepton edms.
For example, at $\tan\beta=10$:\\
we have ${f}$ at $2.5\times 10^{15}$ GeV
\begin{eqnarray} {f}&=&Diag[\left(\matrix{
  6.4\times10^{-5}                & 1.6\times 10^{-3}   &1 }\right)]
\end{eqnarray}
The Dirac neutrino coupling is \begin{eqnarray} {Y_{\nu}}&=&\left(\matrix{
  0                & 1.67\times 10^{-2}-1.43\times 10^{-2}i   
  &-2.31\times 10^{-2}-7.7\times 10^{-3}i
\cr
1.67\times 10^{-3}-1.43\times 10^{-2}i&6.4\times 10^{-2}+1.1\times
10^{-1}i &8.3\times 10^{-2}-1.14\times 10^{-1}i\cr
    -2.31\times 10^{-2}-7.7\times 10^{-3}i&8.3\times 10^{-2}-1.14\times
10^{-1}i &-0.57 i   }\right)
\end{eqnarray} Using these parameters, we find
$\sin^22\theta_{\odot}=0.83$, $\Delta m^2_{\odot}=5\times 10^{-5}$ eV$^2$
and
$\sin^22\theta_{A}=0.88$, $\Delta m^2_A=2\times 10^{-2}$ eV$^2$.
The baryon to photon ratio is $\eta_B\sim6\times 10^{-10}$.

In Fig.3 we plot the electric dipole moments of electron and the muon as
functions of $A_0$ and $m_{1/2}$ and $m_0$ for $\tan\beta=10$. 
The relic density constraint
is satisfied in the only available stau-neutralino co-annihilation region
as before. The magnitude $d_{mu}$ can be as large as $10^{-28.5}$ ecm for
$A_0=800 GeV$, where
as the $d_{e}$ can be as large as $10^{-32}$ ecm. 

In Fig.4, we increase the scale where the universality and
proportionality assumptions are made to
$10^{18}$ GeV and study its effect on $d_{mu}$  
as a function of $m_{1/2}$ for $A_0=300$ GeV. One needs to be careful
about the size of the couplings at this
new scale. The effects of the $f$ couplings in this new region  increase 
the dipole moments. The top line shows the dipole moment for the
increased scale and we conclude that
the edm increases as the scale moves up.

Note that all these values are considerably above the values in the
absence of the $f$ couplings\cite{ellis}. Also note the violation of the
mass scaling law for the dipole moments in the presence of the $f$
couplings.

 \begin{figure}\vspace{0cm}
    \begin{center}
    \leavevmode
    \epsfysize=8.0cm
    \epsffile[75 160 575 630]{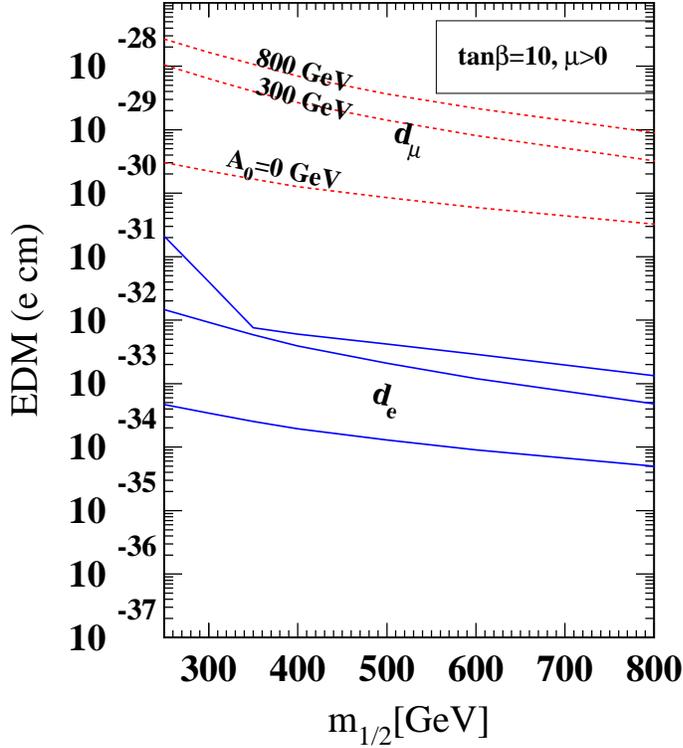}
    %\epsffile[75 160 575 630]{/home/duttabh/lfv/muegmh30.eps}
    \vspace{2.0cm}
     \caption{\label{fig:fig3} Electric dipole moments of electron
$|d_e|$ (solid
     line) and muon $|d_{\mu}|$ (dotted line) for different values of $A_0$ for
     $\tan\beta=10$.}
\end{center}\end{figure}
\begin{figure}\vspace{0cm}
    \begin{center}
    \leavevmode
    \epsfysize=8.0cm
    %\epsffile[75 160 575 630]{edmbdmblel10p.eps}
    %\epsffile[75 160 575 630]{/home/duttabh/lfv/muegmh30.eps}
    \epsffile[75 160 575 630]{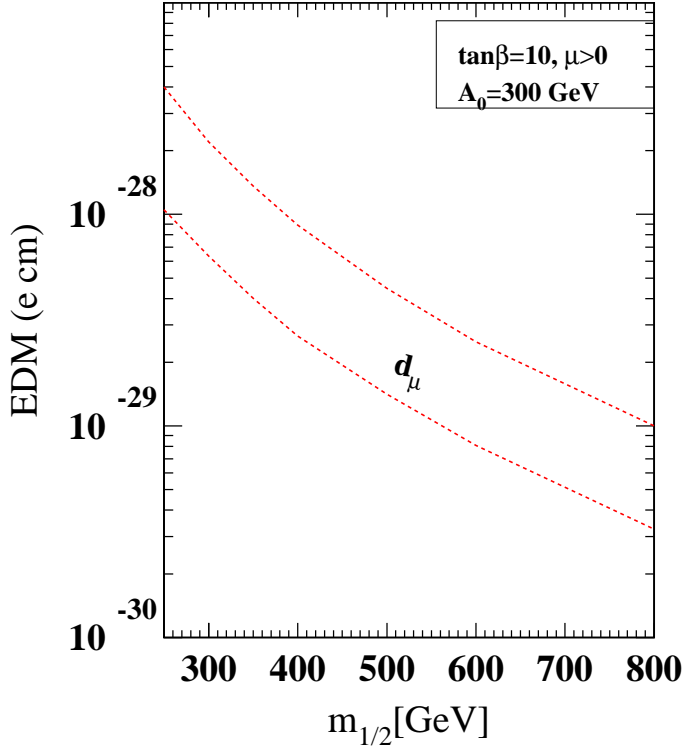}
    \vspace{2.0cm}
     \caption{\label{fig:fig4} Electric dipole moments of muon $d_{\mu}$
 for $M_G=2.3\times 10^{16}$ GeV and $10^{18}$ GeV}
\end{center}\end{figure}

\section{The case of two $\nu_R$'s and the $3\times 2$ seesaw}

In this section, we consider the predictions of a class of models with 
$3\times 2$ seesaw considered in various papers\cite{kuchi,endoh,king}.
First point to note is that the general results we discussed for the case
of three right handed neutrinos in sec.2 do not apply to this case. In
particular, as has been shown by Endoh et al\cite{endoh}, the leptogenesis
phase and the PMNS phases are directly related in this model unlike the
three RH neutrino case.

To proceed with the investigation of this model, one can choose a basis
such that $Y_\ell$ and $f$ are diagonal. The most general $3\times 2$
matrix $Y_\nu$ can now be parameterized as:
\begin{eqnarray}
Y_\nu~=~V\left(\matrix{0 & m_2 & 0\cr 0 & 0 & m_3}\right) U
\end{eqnarray}
where $V~=~\left(\matrix{c&s\cr -s & c}\right)\left(\matrix{e^{i\gamma_R} &
0\cr 0 & e^{-i\gamma_R}}\right)$ and
$U~=~O_{23}(\theta_{23})U_{13}(\theta_{13}, \delta)O_{12}(\theta_{12})P_L$
where $P_L=\left(\matrix{e^{i\gamma_L} & 0\cr 0 & e^{-i\gamma_L}}\right)$.

	In order to calculate the edms and $\eta_B$, we 
will work with the horizontal model of Ref.\cite{kuchi,dm}, which leads to
a specific realization of the $3\times 2$ seesaw formula. The model is
based on the gauge group $ SU(3)_c\times SU(2)_L\times U(1)_Y \times
SU(2)_H$ with fermion assignments given as follows:

\begin{center}
{\bf Table I}

\begin{tabular}{|c||c|}\hline
$\Psi \equiv (L_e, L_{\mu})$ & (1,2,-1,2)\\ \hline
$L_{\tau}$ & (1,2,-1,1) \\ \hline
$E^c \equiv (\mu^c, -e^c)$ & (1,1,-2, 2)\\ \hline
$\tau^c$ & (1,1,-2, 1)\\ \hline
$N^c\equiv (\nu^c_{\mu}, -\nu^c_{e})$ & (1,1,0,2)\\ \hline
$\nu^c_{\tau}$& (1,1,0,1)\\ \hline
$\chi_H\equiv \left(\begin{array}{cc} \chi_{1} & \chi_2 \\
\end{array}\right) $& (1, 1, 0, 2)\\ \hline
$\bar{\chi}_H\equiv \left(\begin{array}{cc} -\bar\chi_{2} & \bar\chi_1 \\
\end{array}\right)$ &
(1,1,0,2)\\\hline
$H_u$ & (1,2,+1,1)\\ \hline
$H_d$ & (1,2,-1,1)\\ \hline
$\Delta_H$ & (1,1,0,3)\\\hline
\end{tabular}
\end{center}

\bigskip

\noindent{\bf Table caption: } We display the quantum number of the matter
and Higgs superfields of our model.

Here $L_{e,\mu,\tau}$ denote the left handed lepton doublet
superfields. Other symbols are self explanatory.
We arrange the Higgs potential in such a way that the $SU(2)_H$ symmetry
is broken by $<\chi_1>=<\bar{\chi}_2>=
v_{H1}; <\chi_2>=<\bar{\chi}_1>=v_{H2}$ and $<\Delta_{H,3}>=
v'_H$, where $v_{H,i},v'_H \gg v_{wk}$. The vevs for $\bar{\chi}$ are
chosen so as
to cancel the D-terms and leave supersymmetry unbroken below the scale
of horizontal symmetry breaking.

The Yukawa superpotential for this model is given by:
 \begin{eqnarray}
 W_Y&=&h_0 (L_eH_u\nu^c_e+L_\mu H_u\nu^c_\mu)
+h_1 L_\tau(\nu^c_\mu \chi_2 + \nu^c_e\chi_1)H_u/M
-if N^{c T}\tau_2{\bf \tau \cdot \Delta_H}N^c \\ \nonumber
&&\frac{h'_1}{M}(L_e\chi_2-L_\mu \chi_1)H_d\tau^c+
\frac{h'_4}{M} L_\tau H_d(\mu^c\chi_2+e^c\chi_1)+
{h'_3}L_{\tau}H_d \tau^c
+h'_2 (L_ee^c+L_\mu \mu^c)H_d
\end{eqnarray}
 The parameters in the above equation
 have been determined to fit neutrino mass data at low
energies\cite{dm}. We use
them in our calculation. We expect the predictions to be
typical of most $3\times 2$ models. From \cite{dm}, we see that at
$\tan\beta=30$, the charged lepton mass matrix is
 given by (at the scale $10^{13}$ GeV):
\begin{eqnarray} {\cal M}_{\ell}&=&\left(\matrix{
  7.5\times 10^{-3} & 0 & -1.135\cr
  0 & 7.5\times 10^{-3} & 0.247\cr
 2.14\times 10^{-2} & 0.0984& 0.85  }\right).
\end{eqnarray}
 This determines all the Yukawa couplings responsible for charged
lepton masses to be $h_1'=1.135\times 10^{-1}$, $h_2'=1.29\times 10^{-3}$,
$h_3'=1.466\times 10^{-1}$ and  $h_4'=9.8\times 10^{-3}$ with
$\kappa_2=10$ and ${\kappa_2\over \kappa_1}=4.588$.
We get  the correct values of charged lepton masses at the weak scale
from the above matrix. We use the MSSM RGEs between the horizontal scale
and the
weak scale. The $f$ is given at the horizontal scale:
\begin{eqnarray} f&=&\left(\matrix{
  5\times 10^{-3} & 0.025+0.499 i\cr
  0.025+0.499 i&2.5\times 10^{-3}  }\right).
\end{eqnarray}
Using the same $\kappa$s as above, $h_0=9.2\times 10^{-2}$ and
$h_1=-2.92\times 10^{-1}$, we find
 $\sin^22\theta_{\odot}=0.87$, $\Delta
m^2_{\odot}=7.5\times 10^{-5}$ eV$^2$,
$\sin^22\theta_{A}=0.88$, $\Delta m^2_A=2.48\times 10^{-3}$ eV$^2$ and 
$|U_{e3}|=0.11$ which are within the experimentally allowed regions.
 We find $\eta_B\sim6\times 10^{-10}$.

In Fig.5 we plot the electric dipole moments of electron and the muon as
functions of
$m_{1/2}$ and $m_0$ for $\tan\beta=30$ and $A_0=300$ GeV. We again
satisfy the relic dark matter density 
constraint. We find that unlike other models, both $d_{mu}$ and $d_e$
are of same order and as large as $10^{-31}$ ecm.

\begin{figure}\vspace{0cm}
    \begin{center}
    \leavevmode
    \epsfysize=8.0cm
    %\epsffile[75 160 575 630]{edm321.eps}
    \epsffile[75 160 575 630]{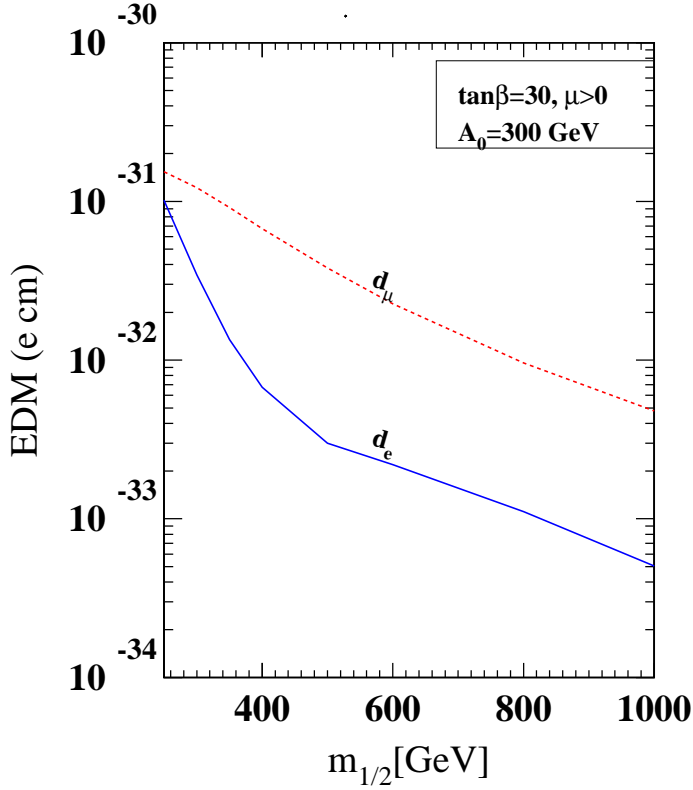}
    %\epsffile[75 160 575 630]{/home/duttabh/lfv/muegmh30.eps}
    \vspace{2.0cm}
     \caption{\label{fig:fig5} Electric dipole moments of electron $d_e$ (solid
     line) and muon $d_{\mu}$ (dotted line) for different values of $A_0$ for
     $\tan\beta=30$. The model parameters are described in the text.}
\end{center}\end{figure}

To be completely phenomenologically consistent, we need to study the
profile of lepton flavor violation for all these models. For model 1,
this has been studied extensively in \cite{babu1} and the results are
within the present bounds but quite testable as has been emphasized.
Similarly, the branching ratio for $\mu\rightarrow e\gamma$ in model 3 
can be found in
ref.\cite{dm}. We will therefore only calculate
BR[$\mu\rightarrow e\gamma$] for model 2 for our choice of parameters and
present them in fig.6 for different values of $m_{1/2}$ and
$A_0$. We again demand the relic density
constraint to be satisfied in this parameter space. We find that the most region of the
parameter space is allowed by this branching ratio. The BR for
$\tau\rightarrow \mu\gamma$ is $\sim
10^{-8}-10^{-14}$ in the same region parameter space.

\begin{figure}\vspace{0cm}
    \begin{center}
    \leavevmode
    \epsfysize=8.0cm
    \epsffile[75 160 575 630]{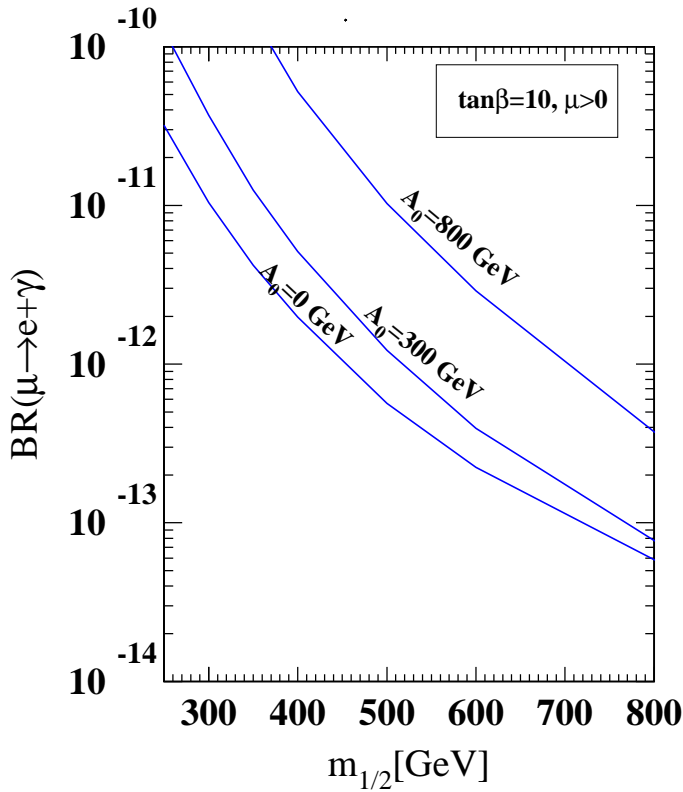}
    %\epsffile[75 160 575 630]{muegmel10.eps}
    %\epsffile[75 160 575 630]{/home/duttabh/lfv/muegmh30.eps}
    \vspace{2.0cm}
     \caption{\label{fig:fig6} The branching ratio for $\mu\rightarrow
e+\gamma$ for model 2.}
\end{center}\end{figure}

\section{Summary and conclusion}
In summary, we have calculated the electric dipole moment of the electron
and the muon in supersymmetric seesaw models where right handed neutrino
masses arise from renormalizable coupling in the superpotential. Using
this
we have attempted to probe the CP violating phases in
the right handed neutrino mass matrix responsible for leptogenesis.
We consider the minimal gauge group which allows for this. We make the
simple and widely used choice for the
supersymmetry breaking parameters as in the mSUGRA models so that no
new phases enter the discussion, other than the phases in the RH
neutrino mass matrix which manifest at low energies and in
leptogenesis. We find
that due to the fact that the right handed neutrino masses
 arise via renormalizable couplings, the lepton edms get
enhanced and in some models come within the range accessible to proposed
experiments. For instance, we can have $d_e$ as large as $10^{-31}$ ecm.
The largest values for muon edm we predict are at the level of
$10^{-27.3}$ ecm. Also, the scaling law $d_mu/d_e~=~m_\mu/m_e$ does not
hold in general.
All the parameters of our model are chosen so that they are
consistent with present neutrino data and
the produce the required baryon asymmetry. The smallness of the edm values
in most cases is due to the assumptions of complete universality of scalar
masses in the MSSM and the proportionality of the $A$ terms to the Yukawa
coupling.  To the extent that collider experiments have the potential to
confirm or rule out the mSUGRA models in the near future, these results
can teach us important things about the origin of matter as well as about
the origin of neutrino masses.

Clearly if the assumtions about SUSY breaking terms is relaxed, one can
enhance the edm
values further. One must then make sure that new phases not connected to
leptogenesis are not responsible for the enhancement. In the minimal model
there is no such confusion. This possibility is presently under
consideration.

The work of R. N. M. is supported by the National Science Foundation
 Grant No. PHY-0099544 and that of B. D. by the Natural Sciences and
Engineering Research Council of Canada. We like to thank K. S. Babu,
S. Davidson and W. Buchmuller for
discussions. One of us (R. N. M.) would like to thank G. Kane and the
Michigan Center for Theoretical Physics for hospitality when the work was
in its last stages and for the opportunity to
present this work at the workshop on {\it Baryogenesis} prior to
publication.

\end{document}